# Cluster Detection Capabilities of the Average Nearest Neighbor Ratio and Ripley's K Function on Areal Data: An Empirical Assessment

Nadeesha Vidanapathirana[1], Yuan Wang[1], Alexander C. McLain[1], Stella Self[1]

[1]Department of Biostatistics and Epidemiology, University of South Carolina, Columbia, SC.

*Spatial clustering detection methods are widely used in many fields including epidemiology, ecology, biology, physics, and sociology. In these fields, areal data is often of interest; such data may result from spatial aggregation (e.g. the number disease cases in a county) or may be inherent attributes of the areal unit as a whole (e.g. the habitat suitability of conserved land parcel). This study aims to assess the performance of two spatial clustering detection methods on areal data: the average nearest neighbor (ANN) ratio and Ripley's K function. These methods are designed for point process data, but their ease of implementation in GIS software (e.g., in ESRI ArcGIS) and the lack of analogous methods for areal data have contributed to their use for areal data. Despite the popularity of applying these methods to areal data, little research has explored their properties in the areal data context. In this paper we conduct a simulation study to evaluate the performance of each method for areal data under various areal structures and types of spatial dependence. These studies find that traditional approach to hypothesis testing using the ANN ratio or Ripley's K function results in inflated empirical type I rates when applied to areal data. We demonstrate that this issue can be remedied for both approaches by using Monte Carlo methods which acknowledge the areal nature of the data to estimate the distribution of the test statistic under the null hypothesis. While such an approach is not currently implemented in ArcGIS, it can be easily done in R using code provided by the authors.*

## 1 Introduction

Researchers have been using spatial clustering analysis for many years to analyze data for spatial patterns. One of the earliest examples of spatial clustering analysis in public health occurred when John Snow mapped the location of cholera cases to identify the source of an outbreak in London in 1894 (Moore & Carpenter, 1999). Over the past few decades, the popularization of geographical information systems (GIS) software has led to the development of many new methods for spatial analysis. A Google Scholar search for 'spatial clustering' returns 15,600 results published from 2001 to 2010, with an increase to 28,000 results published from 2011 to 2020. Both point process (latitude-longitude) and areal data can exhibit spatial patterns, and a variety of methods have been developed to analyze spatial patterns in both types of data.

One of the oldest techniques for detecting spatial clustering is the average nearest neighbor (ANN) ratio (Clark & Evans, 1954). Given a set of 'observation locations' (e.g. coordinate locations of a specific plant species, street addresses of violent crimes, etc.), the ANN method computes the distance between each observation location and its nearest neighbor and uses the average 'nearest neighbor distance' to compute a test statistic. The ANN ratio has been widely used to detect clustering in various types of point process data, including disease cases (Aziz et al., 2012; Khademi et al., 2016; Melyantono, Susetya, Widayani, Tenaya, & Hartawan, 2021), crime hotspots (Brookman-Amissah, Wemegah, & Okyere, 2014; Wing & Tynon, 2006; Z. Zhang et al., 2020) and archeological artifacts (Kıroğlu, 2003; Whallon, 1974). The ANN ratio has also been used to detect clustering in areal data (after mapping to centroids) in disease case data (Mollalo, Alimohammadi, Shirzadi, & Malek, 2015) aggregated to the municipality level.

Ripley's K function was first developed in 1976 and can simultaneously describe spatial patterns at different scales. For a given distance $t$, Ripley's K function returns the expected number of observation locations within a distance of $t$ from a randomly selected observation location.

Ripley's K function can be used to determine if data are clustered (or dispersed) at a specific geographic scale determined by $t$. Ripley's K function has been used to assess point process data for clustering in a variety of fields including epidemiology (Hohl, Delmelle, Tang, & Casas, 2016; Lentz, Blackburn, & Curtis, 2011; Ramis et al., 2015), ecology (Haase, 1995; Moeur, 1993; Wolf, 2005), and sociology (Lu & Chen, 2007; Vadlamani & Hashemi, 2020). It has also been applied to areal data (after mapping the areal units to their centroids) in a variety of applications including disease case data aggregated at the municipality level (Karunaweera et al., 2020; Mollalo et al., 2015; Skog, Linde, Palmgren, Hauska, & Elgh, 2014), locations of conserved land (Zipp, Lewis, & Provencher, 2017), locations of land parcels in industrial use (Qiao, Huang, & Tian, 2019), human-wildlife interactions (Kretser, Sullivan, & Knuth, 2008), and rockfall events (Tonini & Abellan, 2014).

The ANN ratio and Ripley's K function were developed for point process data. However, Environmental Systems Research Institute (ESRI) ArcGIS software allows the user to implement both these methods on areal data using their Average Nearest Neighbors Tool (ANN ratio) and the Multidistance Spatial Cluster Analysis Tool (Ripley's K function); both tools are in the Spatial Statistics Toolbox. When the user applies either method to areal data (i.e. polygon features), ArcGIS automatically maps each polygon feature to its centroid and applies the method to the resulting set of points (Esri, 2021a; Esri, 2021b). To our knowledge, the performance of the ANN ratio and Ripley's K function under these conditions has never been evaluated. Both methods assume that under the null hypothesis, the observation locations arise from a homogenous Poisson process. However, for areal data the centroids of smaller units tend to be closer to the centroids of their neighbors when compared to the centroids of larger units, making it improbable that the locations of the centroids will follow a homogeneous Poisson process even in the absence of spatial clustering. As the homogeneous Poisson process assumption is likely violated for most areal data structures, it is advisable to assess the performance (i.e. empirical type I error rate and empirical power) of the ANN ratio and Ripley's K function when applied to areal data.

In this paper, we conduct a simulation study to evaluate the performance of each method as implemented in ArcGIS for areal data under different types of spatial dependence and areal structures. We compare the performance of each method under the standard ArcGIS implementation with an alternate implementation in which the null distribution of each test statistic is estimated using Monte Carlo methods that acknowledge the underlying areal structure of the data. Section 2 presents a detailed description of the two spatial clustering methods. Section 3 describes the simulation study and Section 4 describes its results. Section 5 contains discussion and concluding remarks.

## 2 Methods

The ANN ratio and Ripley's K function can be used to perform a hypothesis test for the presence of spatial clustering or dispersion. Traditionally, the null hypothesis for each of these tests is that the observed locations exhibit complete spatial randomness (CSR), that is, the observation locations arise from a two-dimensional homogeneous Poisson process. Formally, a stochastic process is said to be a homogeneous Poisson process with rate $\lambda$ if the number of events in any bounded region $A$, denoted $N(A)$, is Poisson distributed with mean intensity $\lambda|A|$, that is, $\Pr(N(A) = n) = e^{-\lambda|A|}(\lambda|A|)^n/n!$, where $|A|$ denotes the area of $A$. Given that there are $n$ events in $A$, those events form an independent random sample from a uniform distribution on $A$ (Cressie, 1994).

**Average Nearest Neighbor Ratio**

The average nearest neighbor (ANN) ratio was initially developed to classify spatial patterns in plant populations (Clark & Evans, 1954). In this context, the observation locations consist of locations of observed plants, measured as coordinates in two-dimensional space. This method quantifies the randomness (or lack thereof) among the observed point locations by measuring the distance from each point to its nearest neighbor and using these distances to compute the average nearest neighbor (ANN) ratio given by

$$R = \frac{\bar{r}_O}{\bar{r}_E} \qquad (1)$$

where $\bar{r}_O = \frac{\sum_{i=1}^{N} r_i}{N}$, $r_i$ denotes the distance from the $i^{th}$ individual to its nearest neighbor, $N$ denotes the total number of observations, $\bar{r}_E = \frac{1}{2\sqrt{\rho}}$ is the expected value of $\bar{r}_O$ under CSR for an infinite study area, $\rho = \frac{N}{|A|}$ is the density of the observed distribution, and $|A|$ is the size of the study area. Under CSR, $E(R) = 1$, and for a perfectly clustered distribution (i.e., all points fall at the same location), $E(R) = 0$. Values of $R$ greater than 1 indicate that the points are distributed more uniformly than expected under CSR, while values less than 1 indicate that the points are clustered. The standard approach to hypothesis testing using the ANN ratio assumes that the distribution of $r_O$ under the null hypothesis of CSR is approximately normal. A z-score for the statistic is calculated by

$$z = \frac{\bar{r}_O - \bar{r}_E}{\sigma_{\bar{r}_E}}, \qquad (2)$$

where $\sigma_{\bar{r}_E}$ is the standard error of the mean distance to the nearest neighbor under CSR. It can be shown that under CSR with an infinite study area, $\sigma_{\bar{r}_E} = \frac{0.26136}{\sqrt{N\rho}}$. A significantly negative z-score indicates clustering, and a significantly positive score indicates dispersion. See Figure 1 for examples of clustering, dispersion, and complete spatial randomness.

Clark and Evans (1954) note some limitations of their procedure, such as the sensitivity to the chosen study area and inability to distinguish between certain types of spatial dependence (e.g., tightly clustered points in one place vs pairs of points scattered in population). In this situation, they suggest an extension to this measure by constructing a circle for each observation with an infinite radius, dividing the circle into equal sectors, and measuring the distance from the individual to its nearest neighbor for each of the sectors. They also point out that problems may arise when the data consist of large areal units rather than points and the centroid of each unit is used to calculate the ANN ratio (Clark & Evans, 1954).

In the areal data context, the observations consist of a set of areal units (e.g. counties, census tracts, etc.), with some subset of these units having some characteristic of interest (e.g. counties which imposed a mask mandate during the COVID-19 pandemic, census tracts deemed to be food deserts, etc.). The 'standard implementation' of ANN-based hypothesis testing for such datasets (i.e. the approach taken by ESRI ArcGIS) is to map each areal unit to its centroid, compute the z-score given in Equation 2, and make a rejection decision by comparing this z-score to the quantiles of a standard normal distribution. We propose an alternate 'areal implementation' which acknowledges the underlying areal unit structure. Under this approach, the null distribution of the ANN ratio z-score is estimated using Monte Carlo simulations where the data for each simulation is generated by randomly selecting $N$ observation locations from among the areal unit centroids. The ANN ratio z-score is computed for each dataset, empirical quantiles of the ANN ratio under the null distribution are calculated, and a decision is made by comparing the ANN ratio z-score from the observed dataset to these empirical quantiles.

**Ripley's K Function**

One of the limitations of the ANN method is its inability to test point patterns at different scales simultaneously (Ripley, 1977). For example, it is possible for data to be clustered at a small scale and clustered at a larger scale (i.e., the clusters are clustered) or clustered at a small scale but dispersed at a large scale (i.e., the clusters occur at somewhat regular intervals). Ripley's $K$ function is a second-order spatial analysis tool (i.e., uses variances of the distances between observations) that can address the issue of scale-dependent spatial patterns. Here we only consider Ripley's $K$ function for univariate spatial patterns in two dimensions, but it can also be extended for multivariate spatial patterns (i.e.: comparing spatial patterns of two species) (Dixon, 2001). Let $n_t(s)$ denote the number of points within radius t at point s. The function $K$ is given by

$$K(t) = \frac{E(n_t(s))}{\lambda},$$

where $\lambda$ is the density (number of points per unit area) and can be estimated as $\hat{\lambda} = \frac{N}{|A|}$, $N$ is the observed number of points and $|A|$ is the size of the study area. The expectation is taken over the point locations. If the points follow a homogenous Poisson process (i.e., exhibit CSR), then $K(t) = \pi t^2$ which is the area of a circle of radius $t$. An approximately unbiased estimator for $K(t)$ was proposed by Ripley (Dixon, 2001; Ripley, 1976):

$$\widehat{K}(t) = \hat{\lambda}^{-1} \sum_i \sum_{j \neq i} w_{ij}^{-1} \frac{I(d_{ij} < t)}{N}, \qquad (3)$$

where $d_{ij}$ is the distance between the $i^{th}$ and $j^{th}$ points, $I(d_{ij} < t)$ is the indicator function with the value of 1 if $d_{ij} < t$ and 0 otherwise, and $w_{ij}^{-1}$ is a weight associated with locations $i$ and $j$ that corrects for edge effects. Correction for edge effects is required if any distance $d_{ij}$ is greater than the distance between point $i$ and the boundary. Because the points outside the boundary are not included in the calculation of $\hat{K}(t)$, edge effects can lead to a biased estimator of $K(t)$. Various authors have proposed different edge corrections, such as buffer zones (Sterner, Ribic, & Schatz, 1986; Szwagrzyk, 1990) or toroidal edge corrections (Ripley, 1979; Upton & Fingleton, 1985). One of the most commonly used edge corrections assigns $w_{ij}$ a value of 1 if the circle centered at point $i$ which passes through point $j$ is entirely inside the study area and assigns $w_{ij}$ equal to the proportion of the circumference of the circle that falls in the study area otherwise (Dixon, 2001).

To test for CSR, the estimator $\hat{L}(t) = [\hat{K}(t)/\pi]^{1/2}$ is sometimes used in practice, and $E(\hat{L}(t)) = t$ under CSR (Ripley, 1979). If the observed value of $K(t)$ is larger than the expected value of $K(t)$ for a given distance, the distribution is more clustered than expected under CSR at that distance. If the observed value of $K(t)$ is smaller than the expected value of $K(t)$, the distribution is more dispersed than expected under CSR at that distance. Usually, the distribution of $\hat{K}(t)$ under the null hypothesis of CSR is estimated via Monte Carlo simulations, and critical values from the simulated distribution are used to define a rejection region for hypothesis testing.

Computing Ripley's K function can be computationally expensive for large datasets, as the number of distances between each pair of points must be computed. If inference is required, these computations must be repeated for each dataset generated in the Monte Carlo simulations. Several computationally efficient methods have been suggested for use on large datasets. Tang, Feng, and Jia (2015) develop a massively parallel algorithm which uses graphical processing units (GPUs) to compute Ripley's K function for large datasets. G. Zhang, Huang, Zhu, and Keel (2016) propose two methods for computing Ripley's K which combines cloud computing with pre-sorting and efficient storage to reduce computation time. Wang et al. (2020) develop a distributed computing algorithm to compute Ripley's K function for large spatial-temporal datasets using Apache Spark. In addition to these methods, a number of software packages exist which are suitable for computing

Ripley's K on smaller datasets, including the Multi-distance Spatial Cluster Analysis Tool in ESRI ArcGIS (Esri. (2021b)), the spatstat package in R (Baddeley & Turner, 2021), and the splancs package in R (Bivand et al., 2017).

The 'standard implementation' of hypothesis testing using Ripley's K for areal datasets (and the approach used by ESRI ArcGIS) is to map each areal unit to its centroid and compute $\hat{K}(t)$ (or sometimes $\hat{L}(t)$) using the resulting set of points. The null distribution of $\hat{K}(t)$ is estimated using Monte Carlo simulations where each dataset consists of $N$ observation locations generated from a continuous uniform distribution on the study area, that is, the locations exhibit CSR. Hypothesis testing proceeds by comparing the $\hat{K}(t)$ value from the original dataset to the empirical quantiles from the Monte Carlo simulations. We propose an alternate 'areal implementation' for performing hypothesis testing using Ripley's K function. Under this implementation, the null distribution is again estimated using Monte Carlo simulations, but observation locations for each dataset are generated by randomly selecting $N$ points from among all the areal unit centroids. The $\hat{K}(t)$ from the original dataset is then compared to the empirical quantiles from the Monte Carlo simulations to make a rejection decision.

## 3 Simulations

A simulation study was conducted to assess the performance the standard and areal implementations of hypothesis testing using ANN ratio and Ripley's K function. In this study, we consider three areal unit structures: structure $A_1$ is a $20 \times 20$ regular grid where the units are of the same size, structure $A_2$ is the United States (US) counties in the states of North Carolina, Tennessee, South Carolina, Georgia, Alabama, and Mississippi where the units (counties) are roughly the same size but irregularly shaped, and structure $A_3$ is the Canadian forward sortation areas (FSAs) in the provinces of Alberta, Saskatchewan, and Manitoba where the units have vastly different sizes. Areal structures $A_1$, $A_2$, and $A_3$ contain $n_1 = 400$, $n_2 = 549$, and $n_3 = 267$ areal units, respectively. Observed data consists of a subset of the areal units presumed to have some characteristic of interest, and data generation consists of selecting this subset of units. We generate data under CSR and under two different types of clustering for two sample sizes. For each scenario, 500 datasets are simulated.

Data is generated by selecting the observed units from among the $n_a$ total units present in areal structure $A_a$. We consider two sample sizes for each areal structure: $N = \lfloor \frac{n_a}{10} \rfloor$ and $N = \lfloor \frac{n_a}{4} \rfloor$, and generate data under the null hypothesis of no spatial pattern (data generation mechanism (DGM) $D_1$) and under two alternative hypotheses: a single large cluster (DGM $D_2$) and multiple smaller clusters (DGM $D_3$). Under DGM $D_1$, which corresponds to no spatial pattern, the observed units are selected via uniform sampling without replacement from among the $n_a$ total units. Under DGM $D_2$, which corresponds to the single cluster, units in a pre-selected region of the study area are sampled with 10 times higher probability than the units in the rest of the study area. The regions of higher probability are shown in grey in Figure 2. DGM $D_3$, which corresponds to multiple clusters, is an iterative process. First, an areal unit is selected via uniform random sampling, and this unit is observed, along with units which are directly adjacent to it. Another areal unit is then sampled without replacement from the remaining unobserved units and is observed along with all adjacent units. This process is continued until $N$ units have been observed. Examples of data generated on each areal structure under DGMs $D_1$, $D_2$ and $D_3$ are shown in Figures 3-5, respectively.

The ANN ratio is calculated by using the centroids of the observed units as the observation locations. The ANN ratio is known to be sensitive to the choice of the study area (often referred to as the window); we consider two different methods for selecting the window: window 1 is the entire study area and window 2 is the smallest possible rectangle that encloses all the observed locations (note that window 2 is dependent on the observed data, while window 1 is not). For the standard implementation of the ANN method, the null distribution of the test statistic is assumed to be a standard normal distribution. For the areal implementation, the null distribution of the ANN ratio is estimated using 999 Monte Carlo simulations. For both implementations, for DGM $D_1$ we perform an $\alpha = 0.05$ level two-tailed test by rejecting the null hypothesis if the test statistic is less than the 0.025 quantile or greater than the 0.975 quantile of the null distribution. For DGMs $D_2$ and $D_3$, we perform a left-tailed test (indicative of clustering) by rejecting the null hypothesis if the test statistic falls below the 0.05 quantile of the estimated null distribution.

Ripley's K function was also evaluated using the centroids of the observed units as the observation locations. For each areal structure, the performance of Ripley's K function was evaluated at a sequence of five radii, with the smallest radius equal to twice the smallest distance between any two unit centroids and the largest radius equal to one-quarter of the width of the study area. For each dataset, Ripley's K function is evaluated at each radius using the Kest function in R (Baddeley & Turner, 2005) with the correction input as "Ripley". Under the standard implementation, the null distribution of Ripley's K function at each radius is approximated with 999 Monte Carlo simulations, where the $N$ observation locations for each simulation are generated from a continuous uniform distribution on the study area. Under the areal implementation, the null distribution of Ripley's K function at each radius is approximated with 999 Monte Carlo simulations where the $N$ observation locations are selected at random from the $n_A$ areal unit centroid locations. For both implementations, for DGM $D_1$ we perform an $\alpha = 0.05$ level two-tailed test for each radius $r$ by rejecting the null hypothesis if $\widehat{K}(t)$ is less than the 0.025 quantile or greater than the 0.975 quantile of the null distribution. For DGMs $D_2$ and $D_3$, we perform a right-tailed test (indicative of clustering) by rejecting the null hypothesis if $\widehat{K}(t)$ exceeds the 0.95 quantile of the Monte Carlo samples.

We assess the performance via empirical type I error rate and empirical power. For simulations under the null hypothesis ($D_1$), we report the global empirical type I error rate for the ANN ratio and the empirical type I error rate at each radius for Ripley's K function. For simulations under alternative hypotheses ($D_2$ and $D_3$), we report the global empirical power for the ANN ratio and the empirical power at each radius for Ripley's K function.

## 4 Results

Table 1 summarizes the empirical type I error rate (i.e., empirical probability of rejecting the null hypothesis of CSR when units are generated under CSR) and empirical power (i.e., the empirical probability of rejecting the null hypothesis of CSR when units are clustered) of the ANN ratio under both implementations. Under the standard implementation, the empirical type I error rate is inflated under all scenarios except the US counties under window 2. The inflation is generally more pronounced for the smaller sample size. Under the areal implementation, the empirical type I error rate is at its nominal level. However, the empirical power of the areal implementation is somewhat low (<50%) for all single cluster scenarios except the US counties under window 1. The empirical power for of the areal implementation under the multiple cluster scenarios is high (>90%) for the regular grid and the US counties, but somewhat lower (50-65%) for the Canadians FSAs. In most scenarios, the empirical power of the areal implementation of the ANN methods is higher for window 1 (the entire study area) than window 2 (the smallest enclosing rectangle).

Table 2 summarizes the empirical type I error and the empirical power of Ripley's K function under both implementations. The empirical type I error rate of the standard implementation is far above its nominal level in all cases except for the Canadian FSAs under the smallest radius and smallest sample size. The empirical type I error rate of the areal implementation is at or near its nominal level for all scenarios, ranging from 0.02 to 0.09. The empirical power of the areal implementation is high for most of the single cluster scenarios, with the notable exception of Canadian FSAs under the smallest radius. The empirical power of the areal implementation for the multiple cluster scenario is quite high (>90%) for all radii and sample sizes for the regular grid. The empirical power of the areal implementation for the multiple cluster scenarios for the other two areal structures is highest for the second smallest radii with a notable decline as the radius increases. Since the multiple clustering scenario consists of small clusters which are randomly dispersed (as opposed to large clusters or clusters of clusters), this is to be expected.

## 5 Discussion and Conclusion

The standard implementation of both the ANN ratio and Ripley's K function exhibits a highly inflated empirical type I error rate for almost all scenarios considered in the simulation study. This phenomenon likely occurs because centroids of the areal units do not arise from a homogeneous Poisson process under the null hypothesis. As a result, the centroids of the observed units exhibit spatial patterns, even when those units are selected with uniform probability. These findings call into question the reliability of the current ArcGIS implementation of ANN ratio and Ripley's K function for areal data.

The areal implementation of both methods, though imperfect, is far more reliable than the standard implementation. The empirical type I error rate of the areal implementation for both the ANN ratio and Ripley's K function maintains its nominal level across all scenarios. The low power of the areal implementation of the ANN method under most of the single clustering scenarios is likely due to the test statistic being an average of nearest neighbor distances. The small nearest neighbor distances for observations in the cluster are offset by the larger neatest neighbor distances between units outside the cluster, resulting in a less extreme test statistic. The low power of the areal implementation of Ripley's K function for the smallest radius for the Canadian FSAs suggests that the method may not reliably detect clustering at small distances if the areal units are vastly different sizes. Recall that the smallest radius was twice the minimum distance between any two centroids. For the Canadian FSAs, only 6 (0.02%) pairs of centroids fell within this distance, while 52 (0.03%) pairs of centroids fell within this distance for the US counties and 1482 (1.9%) pairs of centroids fell within this distance for the regular grid. The power to detect multiple clustering at this scale appears to be inversely related to the proportion of centroid pairs falling within the specified radius.

These findings have important implications for the use of these methods on areal data. An inflated type I error rate implies a high false discovery rate, meaning that researchers who have applied these methods to areal data may have wrongly concluded that their data exhibited a spatial pattern when in fact it does not. In epidemiology, this may manifest as the discovery of non-existent disease clusters or outbreaks. In ecology, this the misapplication of these methods may lead researcher to conclude that parcels of conserved land are clustering together to create larger, higher quality habitats, when in fact they are not. In civil engineering and urban planning, the use of these methods could cause researcher to conclude that parcels of land designated for certain uses (parking, recreation, open space, etc.) are well dispersed throughout an urban area when such is not the case. As ESRI ArcGIS software automatically applies the standard implementation of these methods to areal data, these results are particularly concerning for ESRI users. R code which performs the areal implementation of both methods is available at https://github.com/scwatson812/Areal_ANN_and_Ripleys_K. The integration of the areal implementation of these methods into ArcGIS and the development of other spatial clustering detection methods which are better suited for areal data is an excellent area for future work.

## Funding

SS and AM were partially supported by National Institutes of Health grant NIGMS P20GM130420.

Hohl, A., Delmelle, E., Tang, W., & Casas, I. (2016). Accelerating the discovery of space-time patterns of infectious diseases using parallel computing. *Spatial and spatio-temporal epidemiology, 19*, 10-20.

Karunaweera, N. D., Ginige, S., Senanayake, S., Silva, H., Manamperi, N., Samaranayake, N., . . . Zhou, G. (2020). Spatial epidemiologic trends and hotspots of Leishmaniasis, Sri Lanka, 2001–2018. *Emerging infectious diseases, 26*(1), 1.

Khademi, N., Reshadat, S., Zanganeh, A., Saeidi, S., Ghasemi, S., & Zakiei, A. (2016). Identifying HIV distribution pattern based on clustering test using GIS software, Kermanshah, Iran. *HIV & AIDS Review, 15*(4), 147-152.

Kıroğlu, M. F. (2003). *A GIS based spatial data analysis in Knidian amphora workshops in Reşadiye.* Citeseer,

Kretser, H. E., Sullivan, P. J., & Knuth, B. A. (2008). Housing density as an indicator of spatial patterns of reported human–wildlife interactions in Northern New York. *Landscape and Urban Planning, 84*(3-4), 282-292.

Lentz, J. A., Blackburn, J. K., & Curtis, A. J. (2011). Evaluating patterns of a white-band disease (WBD) outbreak in Acropora palmata using spatial analysis: a comparison of transect and colony clustering. *PloS one, 6*(7), e21830.

Lu, Y., & Chen, X. (2007). On the false alarm of planar K-function when analyzing urban crime distributed along streets. *Social science research, 36*(2), 611-632.

Melyantono, S. E., Susetya, H., Widayani, P., Tenaya, I. W. M., & Hartawan, D. H. W. (2021). The rabies distribution pattern on dogs using average nearest neighbor analysis approach in the Karangasem District, Bali, Indonesia, in 2019. *Veterinary World, 14*(3), 614.
17

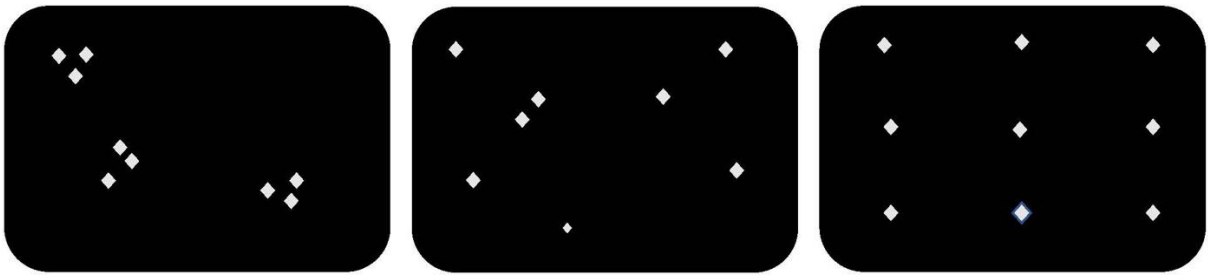

Figure 1. Examples of spatial patterns of clustering (left), CSR (middle), and dispersion (right).

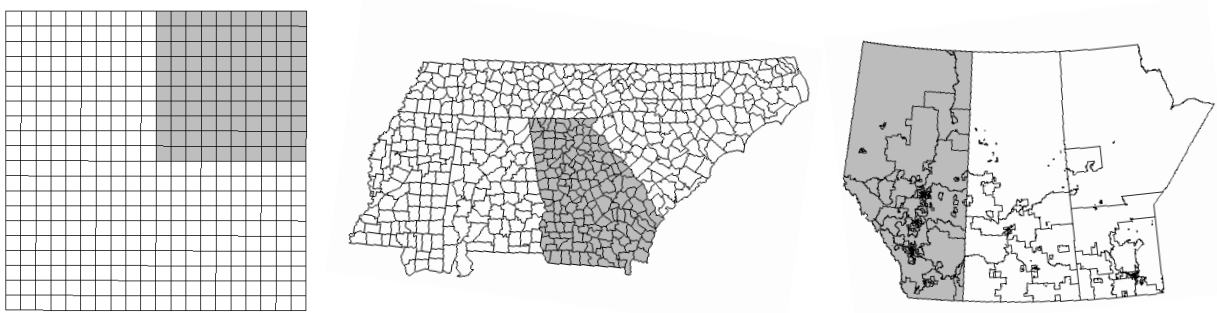

Figure 2. The regions of higher probability selected (grey) under single cluster (left 20x20 regular grid, middle USA counties in six states and right CA FSAs in three provinces).

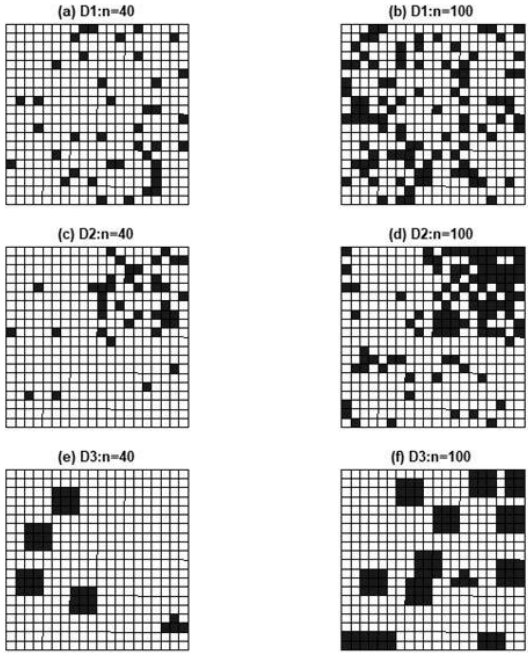

Figure **Error! No text of specified style in document.**. Examples datasets generated for the regular grid under DGMs 1 (top), 2 (middle) and 3 (bottom) for the smaller sample size (left) and larger sample size (right).

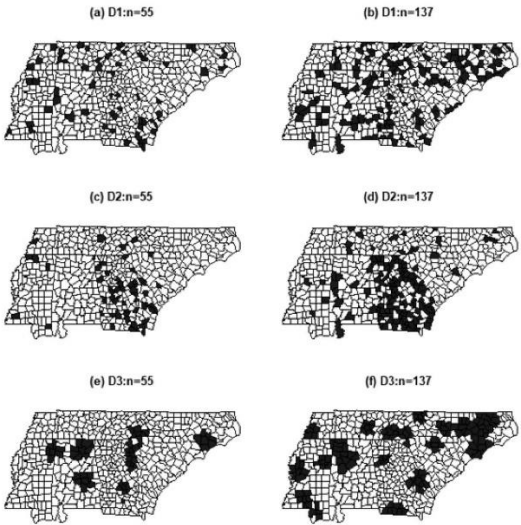

Figure 4. Examples datasets generated for the US counties under DGMs 1 (top), 2 (middle) and 3 (bottom) for the smaller sample size (left) and larger sample size (right).

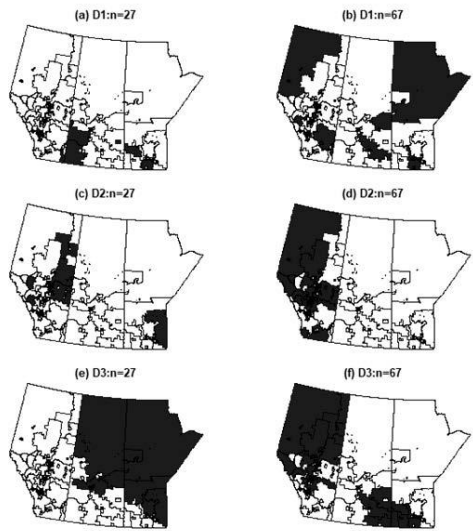

Figure 5. Examples datasets generated for the Canadians FSAs under DGMs 1 (top), 2 (middle) and 3 (bottom) for the smaller sample size (left) and larger sample size (right).

Table 1. Empirical type I error and empirical power of the ANN ratio. $n_a = 400$ for regular grid, $n_a = 549$ for USA counties, $n_a = 267$ for Canadian FSAs.

| Window | | Empirical type I error/ empirical power | | | |
|---|---|---|---|---|---|
| | | Window 1 | | Window 2 | |
| | N | $\lfloor \frac{n_a}{10} \rfloor$ | $\lfloor \frac{n_a}{4} \rfloor$ | $\lfloor \frac{n_a}{10} \rfloor$ | $\lfloor \frac{n_a}{4} \rfloor$ |
| DGM | | The regular grid ($A_1$) | | | |
| $H_o$: CSR: $D_1$ | S | 1.00 | 0.39 | 1.00 | 0.75 |
| | A | 0.05 | 0.05 | 0.05 | 0.04 |
| $H_a$: Single cluster: $D_2$ | S | 0.00 | 0.04 | 0.00 | 0.00 |
| | A | 0.34 | 0.12 | 0.18 | 0.08 |
| $H_a$: Multiple clusters: $D_3$ | S | 0.00 | 1.00 | 0.00 | 0.53 |
| | A | 1.00 | 1.00 | 0.92 | 0.98 |
| | | The US counties ($A_2$) | | | |
| $H_o$: CSR: $D_1$ | S | 0.99 | 0.35 | 0.01 | 0.06 |
| | A | 0.07 | 0.05 | 0.06 | 0.04 |
| $H_a$: Single cluster: $D_2$ | S | 0.00 | 0.27 | 0.36 | 0.24 |
| | A | 0.76 | 0.92 | 0.19 | 0.47 |
| $H_a$: Multiple clusters: $D_3$ | S | 0.95 | 1.00 | 0.96 | 0.99 |
| | A | 1.00 | 1.00 | 0.99 | 0.98 |
| | | The Canadian FSAs ($A_3$) | | | |
| $H_o$: CSR: $D_1$ | S | 1.00 | 0.99 | 1.00 | 0.83 |
| | A | 0.06 | 0.05 | 0.05 | 0.06 |
| $H_a$: Single cluster: $D_2$ | S | 1.00 | 1.00 | 1.00 | 0.76 |
| | A | 0.07 | 0.09 | 0.04 | 0.11 |
| $H_a$: Multiple clusters: $D_3$ | S | 1.00 | 1.00 | 1.00 | 0.93 |
| | A | 0.65 | 0.64 | 0.52 | 0.58 |

S: Standard implementation
A: Areal implementation

Table 2. Empirical type I error rate and empirical power of Ripley's K function.

| | | Empirical type I error/ empirical power | | | | | | | | | |
|---|---|---|---|---|---|---|---|---|---|---|---|
| Radius | | $R_1$ | | $R_2$ | | $R_3$ | | $R_4$ | | $R_5$ | |
| | N | $\lfloor\frac{n_a}{10}\rfloor$ | $\lfloor\frac{n_a}{4}\rfloor$ | $\lfloor\frac{n_a}{10}\rfloor$ | $\lfloor\frac{n_a}{4}\rfloor$ | $\lfloor\frac{n_a}{10}\rfloor$ | $\lfloor\frac{n_a}{4}\rfloor$ | $\lfloor\frac{n_a}{10}\rfloor$ | $\lfloor\frac{n_a}{4}\rfloor$ | $\lfloor\frac{n_a}{10}\rfloor$ | $\lfloor\frac{n_a}{4}\rfloor$ |
| DGM | | The regular grid ($A_1$) | | | | | | | | | |
| $H_o$: CSR: $D_1$ | S | 0.98 | 1.00 | 0.91 | 0.99 | 0.81 | 0.76 | 0.80 | 0.50 | 0.90 | 0.99 |
| | A | 0.08 | 0.03 | 0.08 | 0.03 | 0.07 | 0.06 | 0.06 | 0.05 | 0.06 | 0.03 |
| $H_a$: Single cluster: $D_2$ | S | 0.88 | 0.96 | 0.99 | 1.00 | 1.00 | 1.00 | 1.00 | 1.00 | 1.00 | 1.00 |
| | A | 0.95 | 1.00 | 1.00 | 1.00 | 1.00 | 1.00 | 1.00 | 1.00 | 1.00 | 1.00 |
| $H_a$: Multiple clusters: $D_3$ | S | 1.00 | 1.00 | 1.00 | 1.00 | 1.00 | 1.00 | 1.00 | 0.99 | 1.00 | 0.79 |
| | A | 1.00 | 1.00 | 1.00 | 1.00 | 1.00 | 1.00 | 1.00 | 0.97 | 0.99 | 0.93 |
| | | The US counties ($A_2$) | | | | | | | | | |
| $H_o$: CSR: $D_1$ | S | 1.00 | 1.00 | 0.82 | 0.63 | 0.80 | 0.72 | 0.73 | 0.67 | 0.68 | 0.58 |
| | A | 0.09 | 0.02 | 0.04 | 0.04 | 0.04 | 0.07 | 0.07 | 0.05 | 0.05 | 0.05 |
| $H_a$: Single cluster: $D_2$ | S | 0.17 | 0.00 | 1.00 | 1.00 | 1.00 | 1.00 | 1.00 | 1.00 | 1.00 | 1.00 |
| | A | 0.65 | 0.98 | 1.00 | 1.00 | 1.00 | 1.00 | 1.00 | 1.00 | 1.00 | 1.00 |
| $H_a$: Multiple clusters: $D_3$ | S | 0.36 | 0.00 | 1.00 | 1.00 | 0.99 | 0.98 | 0.86 | 0.85 | 0.74 | 0.69 |
| | A | 0.77 | 0.76 | 1.00 | 1.00 | 0.95 | 0.85 | 0.68 | 0.55 | 0.49 | 0.42 |
| | | The Canadian FSAs ($A_3$) | | | | | | | | | |
| $H_o$: CSR: $D_1$ | S | 0.06 | 0.35 | 1.00 | 1.00 | 1.00 | 1.00 | 0.98 | 1.00 | 0.87 | 0.91 |
| | A | 0.05 | 0.00 | 0.06 | 0.05 | 0.03 | 0.04 | 0.06 | 0.05 | 0.05 | 0.06 |
| $H_a$: Single cluster: $D_2$ | S | 0.35 | 0.31 | 1.00 | 1.00 | 1.00 | 1.00 | 1.00 | 1.00 | 1.00 | 1.00 |
| | A | 0.06 | 0.01 | 0.68 | 0.97 | 0.93 | 1.00 | 0.98 | 1.00 | 1.00 | 1.00 |
| $H_a$: Multiple clusters: $D_3$ | S | 0.53 | 0.50 | 1.00 | 1.00 | 1.00 | 1.00 | 1.00 | 1.00 | 0.83 | 0.80 |
| | A | 0.19 | 0.06 | 0.74 | 0.61 | 0.66 | 0.55 | 0.44 | 0.39 | 0.34 | 0.30 |

S: Standard implementation
A: Areal implementation